\documentclass[12pt]{iopart}
\usepackage{iopams}
\usepackage{setstack}
\usepackage{graphicx}
\begin{document}

\title[Actin network curvature]{Curvature and torsion in growing actin networks}

\author{Joshua W. Shaevitz}
\address{Department of Integrative Biology, University of California, Berkeley, Berkeley, CA. 
        Current Address: Department of Physics and Lewis-Sigler Institute for Integrative Genomics, 
	Princeton University, Princeton, NJ }
\ead{shaevitz@princeton.edu}

\author{Daniel A. Fletcher}
\address{Department of Bioengineering and Biophysics Program, 
	University of California, Berkeley, Berkeley, CA }
\ead{fletch@berkeley.edu}

\begin{abstract}
Intracellular pathogens such as \textit{Listeria monocytogenes} and \textit{Rickettsia rickettsii} move within a host cell by polymerizing a comet--tail of actin fibers that ultimately pushes the cell forward. This dense network of cross--linked actin polymers typically exhibits a striking curvature that causes bacteria to move in gently looping paths. Theoretically, tail curvature has been linked to details of motility by considering force and torque balances from a finite number of polymerizing filaments. Here we track beads coated with a prokaryotic activator of actin polymerization in three dimensions to directly quantify the curvature and torsion of bead motility paths. We find that bead paths are more likely to have low rather than high curvature at any given time. Furthermore, path curvature changes very slowly in time, with an autocorrelation decay time of 200 seconds. Paths with a small radius of curvature, therefore, remain so for an extended period resulting in loops when confined to two dimensions. When allowed to explore a 3D space, path loops are less evident.  Finally, we quantify the torsion in the bead paths and show that beads do not exhibit a significant left-- or right--handed bias to their motion in 3D. These results suggest that paths of actin--propelled objects may be attributed to slow changes in curvature rather than a fixed torque.
\end{abstract}

\vspace{2pc}
\noindent{\it Keywords}: actin, motility, network, curvature, torsion
\\
\submitto{\PB}
\maketitle

\section*{Introduction}

The polymerization of cross--linked networks of actin filaments is an important biological phenomenon that drives such diverse processes as cell crawling, phagocytosis and morphogenesis \cite{Wel2002,Pol2003,Fle2004}. These complex networks are organized by a number of proteins, including nucleation-promoting factors that trigger the formation of nascent filaments off the side of existing network filaments \cite{Wel2002,Pol2003,Ger2000a}. Bacteria such as \textit{L. monocytogenes} and \textit{R. rickettsii} propel themselves during infection by expressing a single protein on their cell surfaces, ActA and RickA respectively, which stimulates actin growth by activating the branching complex Arp2/3 \cite{Jen2004}. Actin polymerization at the cell's rear gives rise to a characteristic comet tail that propels the cell through the host cytoplasm. Several biophysical models have been developed to describe the connection between network growth and the generation of force \cite{Mog2003,Ger2000,Car2003,Soo2005}. Despite the extensive theoretical and experimental investigation of polymerization forces and velocities \cite{Mar2004,Par2005}, less is understood about the changes in direction of network growth.

Comet tails are often highly curved and can propel cells and beads in nearly circular or looping paths \cite{Cam2000,Wie2003,Soo2005}. Phenomenological description of the shapes of the comet tails and the trajectories of actin--propelled objects has been used to guage the effect of biophysical and biochemical perturbations on the process of network formation and maintenance \cite{Cam2004,Lac2004,Aue2003}. In recent work that examined moving \textit{Listeria monocytogenes} cells confined to two dimensions, Shenoy \textit{et al.} postulate that a roughly constant torque drives cells in looping paths\cite{She2007}. A detailed quantification of actin--propelled trajectories in three dimensions can yield additional information about all the forces and torques that, on average, act on the moving object. Indeed, theoretical treatment of network curvature has been used to place limits on the number and placement of pushing filaments at a load surface \cite{Rut2001}. However, the lack of three-dimensional data on this subject has hindered progress in understanding the mechanisms controlling curvature.

Traditional measurements of object trajectories face two main difficulties. First, experiments often track the in--plane objects movements or fluorescence of actin in the network, inherently 2D techniques that do not allow one to monitor the full 3D complexity of the trajectories. Second, due to the limited depth of field of conventional light microscopy, a micron--sized gap between a coverslip and microscope slide is used to restrict out--of--plane motion, leading to pseudo--2D observations \cite{Cam2004,Jen2004,Wie2003,Cam1999}. Motivated by these limitations and an interest in the mechanism governing movement direction, we sought to measure the 3D trajectories explored by actin--propelled beads in large chambers using three-dimensional laser tracking.

\section*{Methods}

\subsection*{Motility Assays}
Experiments were performed as described previously \cite{Sha}. Carboxylated polystyrene beads 792~$\pm$~23 nm in diameter were coated with the nucleation factor RickA purified from \textit{Rickettsia rickettsii} (a gift from Matthew Welch, University of California, Berkeley). The assay mixture, containing beads in \textit{Xenopus laevis} egg cytoplasmic extract supplemented with an ATP-regenerating mix and Rhodamine-labeled actin, was added to a microscope slide immediately after preparation. We used chambers of two thicknesses to monitor bead motion in pseudo--2D and --3D environments. Approximately 2~$\mu$m chambers were created using 2.1~$\pm$~0.1~$\mu$m polystyrene beads as spacers between a glass slide and coverslip. 80~$\mu$m thick chambers were constructed by separating a microscope slide and coverslip with double-sided tape. Beads moved with a slightly slower velocity, $v$, in the 2D environment than in 3D (Table~\ref{TableParams}), most likely due to a viscous interaction with the top and bottom glass surfaces that affect the Brownian Ratchet mechanism of actin--based motility \cite{Sha}.

\begin{table}
\centering
\caption{Motility parameters for different geometries \label{TableParams}} 
\begin{tabular}{ccccc}
$L$ [$\mu$m] & $v$ [nm s$^{-1}$]$^{\ast}$ & $\kappa_{RMS}$ [$\mu$m$^{-1}$]$^{\dagger}$ & $\zeta$ [$\mu$m]$^{\dagger}$ & $\tau$ [$\mu$m$^{-1}$]$^{\dagger}$\cr 
\hline 
2 & 58 $\pm$ 38 & 0.22 $\pm$ 0.04 & 11 $\pm$ 6 & --- \cr
80 &  76 $\pm$ 10 & 0.16 $\pm$ 0.03 & 16 $\pm$ 9 & -0.02 $\pm$ 0.03 \cr
\hline
$^{\ast} \pm SD$; $^{\dagger} \pm$ SEM
\end{tabular}
\end{table} 

\subsection*{Three-dimensional Laser Tracking}
Our instrument uses a single 809--nm wavelength diode laser to measure microsphere position using a position-sensitive detector that monitors scattered light in the back focal plane \cite{Flo1998,Lan2002}. We use a nano--positioning stage capable of moving the sample in three-dimensions over hundreds of microns at a bandwidth of approximately 100 Hz. In addition, we record fluorescence images concurrent with optical tracking to confirm that bead motion corresponds to network growth.

During an experiment, a feedback routine, implemented in software, allows us to track actin--propelled beads over very long distances. 3D photodiode voltage signals \cite{Lan2002,Pet1998,Pra1999} are sampled at 2~kHz and converted into 3D position in realtime using a 5th order 3D polynomial\cite{Lan2002}. The measured bead coordinates are then saved to disk and used to update the position of the microscope stage such that the microsphere remains in the center of the laser focus. Very low laser powers were used to minimize external loads that might perturb actin network dynamics. These powers yielded typical forces of less than 0.01~fN.

\subsection*{Data Analysis}
Time derivatives of the $x$--, $y$--  and $z$-- position versus time records were performing using a 4th order Savitsky-Golay filter with a time constant of one second\cite{Neu2003}. These quantities were then used to calculate the curvature and torsion as defined by equations \ref{EqCurvDef}, \ref{EqTorsionDef}.

\section*{Results and Discussion}

\begin{figure}
	\centering
	\vspace*{.05in} 
 	\includegraphics[width=8.7cm]{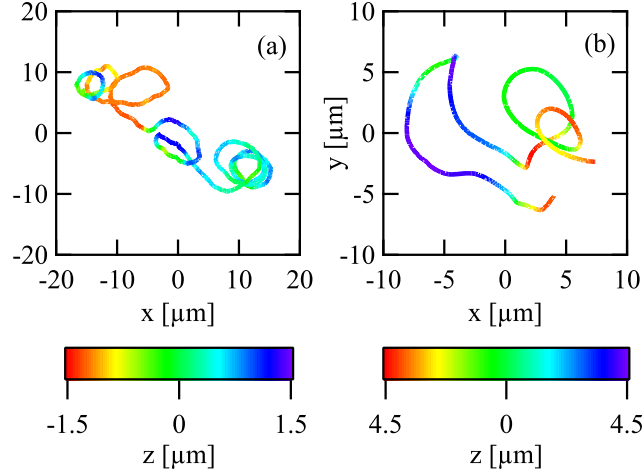}
  	\caption{Example trajectories of 3D bead motion. Data is shown for a bead moving in the 2--$\mu m$ (a) and 80--$\mu m$ (b) chambers. The color scale denotes the $z$--position, the position along the microscope optical axis. \label{FigTraces}}
\end{figure}

We used a variant of 3D optical--force microscopy to track 0.8--$\mu$m diameter beads coated with the nucleation--promotion factor RickA moving in \textit{Xenopus laevis} egg cytoplasmic extract. Using 2--$\mu$m and 80--$\mu$m thick chambers, we created pseudo--2D and --3D environments  to directly test the effect of the reduction of dimensionality. In the 2D chambers ($n=13$), beads proceeded in characteristically looping paths similar to previous observations (Figure \ref{FigTraces}a) \cite{Cam2000,Wie2003}. We observed cases of nearly circular motion as well as shapes which more closely resembled a figure--eight. These paths were markedly different from those we observed using the 3D chambers ($n=15$), where regular looping behavior was not seen (Figure \ref{FigTraces}b). Instead, beads travelled in randomly curved paths, rarely resembling a simple circle or helix.

Using 3D optical tracking, we sought to quantify the shape of the measured bead trajectories. Up to an arbitrary translation and rotation, a geometric path in space can be fully described by two generalized curvatures, termed the curvature and torsion. The curvature, $\kappa$, measures the deviance of a the trajectory from a straight line, where $1/\kappa$ represents the radius of curvature at a given point. Using the coordinates of the trajectory with time, $r(t)$, the curvature can be written as 

\begin{equation}
  \kappa = \frac{\vert \dot{\mathbf{r}} \times \ddot\mathbf{r} \vert}{\vert \dot{\mathbf{r}} \vert ^{3}}
  \label{EqCurvDef}
\end{equation}

The torsion, $\tau$, measures the deviance of the trajectory from a planar curve and can be written as

\begin{equation}
 \tau = \frac{\dot{\mathbf{r}} \cdot \left( \ddot{\mathbf{r}} \times \dddot{\mathbf{r}} \right) }{\vert \dot{\mathbf{r}} \times \ddot{\mathbf{r}} \vert ^2 }
  \label{EqTorsionDef}
\end{equation}
If the torsion is zero, the trajectory lies completely in one plane. Conversely, a positive torsion corresponds to a right--handed helicity whereas a path with negative torsion has a left--handed helical component.

We measured the curvature for beads moving in either two or three dimensions using chambers of two different thicknesses. The RMS curvature measured for these two situations, $\sim 0.2 \mu m^{-1}$ (Table~\ref{TableParams}), are within error of each other and are in good agreement with the curvature deduced from previous video--microscopy studies of motion of both beads and bacteria of approximately $0.1 \mu m^{-1}$  \cite{Cam2004,Wie2003,Soo2005}.

\begin{figure}
	\centering
	\vspace*{.05in} 
  	\includegraphics{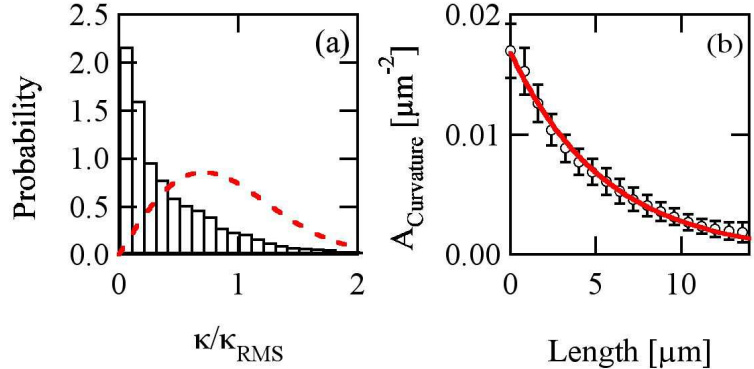}
  	\caption{(a) The curvature probability distribution. The probability of measured curvatures for all beads using the 80--$\mu m$ chambers is displayed (bars). The curvature has been normalized by its RMS value from each run. For comparison,the distribution predicted by Rutenberg and Grant \cite{Rut2001} is shown (red dotted line). (b) Autocorrelation of the path curvature. The autocorrelation of the curvature as a function of path length for a single bead trajectory (open circles) and a fit of this data to a single exponential decay (red solid line) \label{FigCurvature}}
\end{figure}

To our knowledge, only one theoretical treatment of actin--driven motion has considered the molecular mechanism that produces looping trajectories. Rutenberg and Grant \cite{Rut2001} showed that a small number of randomly distributed actin filaments pushing a load will give rise to an intrinsic curvature due to the non--zero instantaneous torque produced by summing the forces from each filament. For a uniformly--coated bead this curvature is given by

\begin{equation}
 \label{EqrmsCurv}
	\kappa _{RMS} = \frac{1}{2a} \sqrt{\frac{3}{2n}}
\end{equation}
where $a$ is the bead radius and $n$ is the number of pushing filaments. In this model, the distribution of curvatures depends only on the RMS value, and for a bulk, 3D geometry can be written as \cite{Rut2001}

\begin{equation}
\label{EqcurvDistribution}
P\left( \kappa \right) = 2\frac{\kappa}{\kappa_{RMS}}e^{- \left(\kappa / \kappa_{RMS} \right) ^2}
\end{equation}
This distribution contains a single broad peak at $\kappa = \kappa_{RMS} / \sqrt{2}$ (Figure \ref{FigCurvature}a dotted line), and thus beads are predicted to move with a ``preferred'' curvature which leads to looping paths.

To directly test this model we calculated the curvature probability distribution from our measured 3D trajectories (Figure~\ref{FigCurvature}a). Rather than a peaked distribution, we find that the curvature distribution decays monotonically, with smaller curvatures, i.e. straighter path segments, more probable that curved ones. Our data appears to rule out the simple model of Rutenberg and Grant.

Bead paths most often had low curvatures; fifty percent of the time the radius of curvature was greater than $25 \mu m$. If this is true, how can we reconsile the appearance of looping trajectories such as those shown in Figure~\ref{FigTraces}a in the 2D chambers? An alternative model to that proposed in reference \cite{Rut2001} is that while smaller curvatures are preferred, network curvature can only change via slow dynamics. If the time to change path curvature is greater than the time to go around a loop, $t_{loop}=2 \pi / \kappa v$, a path will necessarily complete a loop before significantly altering its trajectory. These dynamics may be associated, for instance, with the time required to significantly restructure the micro--architecture of the filaments.

We quantified the temporal dynamics of trajectory curvature by calculating the autocorrelation of the measured curvatures over time (Figure~\ref{FigCurvature}b). The autocorrelation decays exponentially with an average decay constant, $\zeta$, of $16 \pm 9 \mu$m for 3D movement, corresponding to an decay time of $t_{decay} \sim 210$~seconds. In the 2D chambers, the decay constant was lower than in 3D, measuring $11 \pm 5 \mu$m, but corresponding to approximately the same decay time, 190~seconds because of the slower measured velocity.

From this data it is possible to describe why beads appear to move in looping paths when confined to 2D. While a path is most likely to be fairly straight at any instance in time, there is a small probability it will become highly curved. If a highly curved path is initiated, the slow curvature dynamics cause the path to remain highly curved for an extended period resulting in loops when confined to two dimensions. When allowed to explore a 3D space, where a constant curvature does not lead to planar loops, highly--curved path sections merely look ``curly'' rather than looping.

What might cause the $200~s$ time scale, which is much slower than the characteristic times for actin monomer addition, $\sim 100 s^{-1}$, or the addition of new filaments to the network,$\sim 10 s^{-1}$ \cite{Mog2003}? While our beads are round and uniformly coated with RickA, there may be some residual asymmetry to the spatial distribution of the nucleator. Assuming that this density affects the shape and dynamics of network formation, a rotation of the bead relative to the network could yield a change in path curvature. Therefore, it is possible that this time may correspond to the rotational diffusion time for a bead attached to a growing actin network. A second explanation could involve architectural changes in the actin network itself. Recent evidence indicates that a growing actin network contains memory of past forces it has encountered \cite{Par2005}. The actin networks propelling the bead may actively remodel their micro--architecture in response to load and other factors. We propose that the $200~s$ relaxation time we observe in the curvature autocorrelation function may represent the time--scale over which actin remodeling occurs.

\begin{figure}
	\centering
	\vspace*{.05in}
	\includegraphics{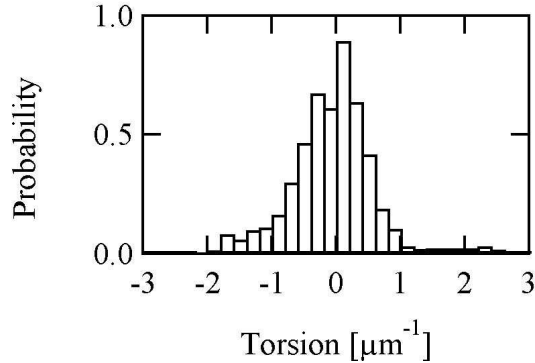}
	\caption{The torsion probability distribution.  The torsion of measured curvatures for all beads using the 80--$\mu m$ chambers is displayed. \label{FigTorsion}}
\end{figure}

In addition to the curvature, we calculated the torsion for beads moving in 3D. The distribution of measured torsions was roughly symetric about zero with a small negative skew (Figure~\ref{FigTorsion}). The average torsion for all 3D runs was $-0.02 \pm 0.03$, not significantly different than zero. Small sections of bead movement often appeared helical (Figure~\ref{FigTraces}b), however overall our data shows that beads do not move in consistently helical paths. In addition, we measured the autocorrelation of the torsion and found no significant correlation over time. The average autocorrelation decayed to zero with a time constant equal to the smoothing time constant used to calculate the derivatives of the position with time, approximately 5 seconds.

Our data is markedly different from the right--handed helical tails which have been observed with \emph{Listeria monocytogenes} moving in high--speed brain extracts \cite{Zei2005}. These trajectories, along with data that moving \emph{Listeria} cells rotate around their long--axis \cite{Rob2003} suggest that our RickA--coated beads and \emph{Listeria} might move by slightly different mechanisms. First, \emph{Listeria} use the nucleation--promoting factor ActA, which might produce a slightly different network architecture than that produced by RickA. Electron-micrographs of \emph{Rickettsia} cells show comet tails with filaments arranged in a parallel geometry \cite{Gou2005}, although beads coated with RickA have been shown to produce dendritic networks similar to those generated by ActA \cite{Jen2004}. Second, \emph{Listeria} cells have been reported to produce a helical surface pattern of ActA \cite{Raf2006}, which may give rise to a network that applies a constant torque to moving cells, producing helical trajectories and the looping trajectories seen in 2D \cite{She2007}. Our uniformly coated bead would not experience this torque.

\section*{Conclusion}
We have presented experimental measurements of the 3D paths explored by actin--propelled beads. In a pseudo--2D environment, beads appear to move in looping paths as has been shown previously for measurements using video--microscopy. In contrast, relaxing the restriction to 2D movement, we find that in 3D beads do not move in looping paths, but still exhibit curved, complex structures. Differing from observations of moving \emph{Listeria} cells, our beads do not move in helical paths in 3D environment, and instead we measure an average torsion of zero.

While this new data places constraints on the physical mechanisms that give rise to the movement of RickA--coated beads, it also speaks to the geometric persistence inherent in an actin network. The correlation time we observe in the curvature of actin--propelled bead paths may be a limiting time for the change in direction of a growing actin network. Biochemical events that attempt to change the direction of a growing actin network, such as chemotaxis or phagocytosis, must do this in the presence of an intrinsic curvature and temporal correlation like that described here.

\newpage
\section*{References}
\bibliographystyle{unsrt}

\end{document}